\documentclass[aps,prl,groupedaddress,showpacs,showkeys]{revtex4-1}

\usepackage{graphicx}
\usepackage{dcolumn}
\usepackage{bm}
 
\def\mbi#1{\mbox{\bfseries\itshape #1}}

\newcommand{\araa}{\rm Ann.~Rev.~Astron.~\&~Astrophys.~}
\newcommand{\aap}{\rm Astron.~\&~Astrophys.~}

\newcommand{\apjl}{\rm Astrophys.~J.~Lett.~}
\newcommand{\apjs}{\rm Astrophys.~J.~Supp.~}

\newcommand{\jcap}{\rm JCAP~}

\newcommand{\mnras}{\rm Mon.~Not.~R.~Astron.~Soc.~}
\newcommand{\na}{\rm New~Astron.~}

\newcommand{\physrep}{\rm Phys.~Rep.~}

\begin{document}


\title{Effects of power law primordial magnetic field on big bang nucleosynthesis}

\author{Dai G. Yamazaki$^{1}$}
 \email{yamazaki.dai@nao.ac.jp}
\author{Motohiko Kusakabe$^{2}$}%
\affiliation{%
$^{1}$National Astronomical Observatory of Japan, Mitaka, Tokyo 181-8588, Japan}%
\affiliation{%
$^{2}$Institute for Cosmic Ray Research, University of Tokyo, Kashiwa, Chiba 277-8582, Japan}%
\date{\today}

\begin{abstract}
Big bang nucleosynthesis (BBN) is affected by the energy density of a primordial magnetic field (PMF).  For an easy derivation of constraints on models for PMF generations, we assume a PMF with a power law (PL) distribution in wave number defined with a field strength, a PL index, and maximum and minimum scales at a generation epoch.  We then show a relation between PL-PMF parameters and the scale invariant (SI) strength of PMF for the first time.  We perform a BBN calculation including PMF effects, and show abundances as a function of baryon to photon ratio $\eta$.  The SI strength of the PMF is constrained from observational constraints on abundances of $^4$He and D. The minimum abundance of $^7$Li/H as a function of $\eta$ slightly moves to a higher $^7$Li/H value at a larger $\eta$ value when a PMF exists during BBN. We then discuss degeneracies between the PL-PMF parameters in the PMF effect.  In addition, we assume a general case in which both the existence and the dissipation of PMF are possible.  It is then found that an upper limit on the SI strength of the PMF can be derived from a constraint on $^4$He abundance, and that a lower limit on the allowed $^7$Li abundance is significantly higher than those observed in metal-poor stars.
\end{abstract}

\pacs{26.35.+c, 98.62.En, 98.80.Es, 98.80.Ft}
\keywords{Big Bang nucleosynthesis, Magnetic field}
\maketitle
\section{\label{sec:introduction}Introduction}
Element abundances in the early Universe are important observables to determine the features of the physical processes at the epoch of big bang nucleosynthesis  (BBN). 
From a discovery of expansion of Universe \cite{1929PNAS...15..168H} and the development of the cosmological theory \cite{1916AnP...354..769E,1924ZPhy...21..326F,1927ASSB...47...49L,1933MNRAS..94..159W,1929PNAS...15..168H}, it was suggested that the early Universes was enough hot and dense enough to allow nuclear fusion to be operative, which is the original idea of BBN \cite{1946PhRv...70..572G,1948PhRv...73..803A,1948PhRv...74.1737A,1949PhRv...75.1089A}. It was assumed that radiative neutron capture reaction generates primordial elements in hot and dense conditions. 
However, neutrons and protons can be converted to each other via weak interactions. Hayashi studied this effect and proposed that BBN started from not a pure neutron gas, but a hybrid gas composed of neutrons and protons as well as radiation \cite{1950PThPh...5..224H}. Hayashi studied routes of $^4$He production and indicated that a relic abundance of $^4$He depends on an initial abundance of the neutron \cite{1950PThPh...5..224H}.  This suggestion and the consideration of relativistic quantum statistics refined the BBN model \cite{1953PhRv...92.1347A}. 
The discovery of the cosmic microwave background (CMB) \cite{1965ApJ...142..419P} indicated that the early Universes was hot and dense, and BBN theory was confirmed.
Production of D, $^3$He, $^4$He, $^7$Li \cite{1967ApJ...148....3W}, all stable isotopes of Li, Be and B \cite{1967PThPh..38.1083S}, and heavier elements up to $^{20}$Ne \cite{1956PThPh..16..613H} have been studied. 
Now the BBN theory includes many important physics, and the numerical precision of the BBN calculations is rather high \cite{1985ApJ...293L..53A,1985ARA&A..23..319B,1999PhRvL..82.4176B,2001ApJ...552L...1B,2002PhRvD..65d3510C,2004ApJ...600..544C,2012ApJ...744..158C,1995Sci...267..192C,2001NewA....6..215C,1982PhRvD..26.2694D,2000NuPhB.568..421E,2000JHEP...09..038E,1996NewA....1...77F,1967ARA&A...5..525F,1995PhRvL..75.3977H,2009PhR...472....1I,1990ApJ...358...47K,1995PhLB..347..347K,1993PhR...229..145M,2011PhLB..701..296M,1981ApJ...246..557O,1990PhLB..236..454O,1996IJMPA..11..409O,1996RPPh...59.1493S,1998RvMP...70..303S,1993ApJS...85..219S,2000PhST...85...12T,2000A&A...360...15V,1973ApJ...179..343W}.

The baryon to photon ratio, $\eta$, is one of the important parameters characterizing the BBN model. Recently, we have constrained $\eta$ from observations of CMB \cite{Spergel:2003cb,2007ApJS..170..377S,2009ApJS..180..306D,2011ApJS..192...16L} and tested various models of BBN using the constrained value of $\eta$ as an input.

Effects of a primordial magnetic field (PMF) on BBN have been studied \cite{1969Natur.222..649O,1970ApJ...160..451M,1969Natur.223..938G,1994PhRvD..49.5006C,1996PhRvD..54.4714C,1995APh.....3...95G,1996PhLB..379...73G,1996PhRvD..54.7207K,1997PhRvD..56.3766K,1999PhRvD..59l3002S,Grasso:2000wj}.
The PMF directly affects rates of weak reactions \cite{1969Natur.222..649O,1970ApJ...160..451M}, and also the cosmic expansion rate through the PMF energy density.  Results of BBN could, therefore, be different from that for the case with no PMF \cite{1969Natur.223..938G,1994PhRvD..49.5006C,1996PhRvD..54.4714C,1995APh.....3...95G,1996PhLB..379...73G,1996PhRvD..54.7207K,1997PhRvD..56.3766K}.
The boost of cosmic expansion by the PMF energy density is the dominant effect of the PMF, and the change of weak reaction rates is subdominant \cite{1996PhRvD..54.4714C,1996PhLB..379...73G,1996PhRvD..54.7207K}.  
An updated result of light elements up to Li in the BBN model including a PMF has been derived based on a network calculation consistently taking account of important effects of PMF \cite{2012arXiv1204.6164K}.

A power law (PL) spectrum is one of the most familiar distributions for the energy densities of various physical phenomenon on cosmological scales, including the PMF(e.g. \cite{Grasso:2000wj,2011PhR...505....1K,2012PhR...517..141Y} and references therein). Thus many researchers have been studying the effect of the PMF with the PL spectrum on various physical processes in the early Universe \cite{Grasso:2000wj,2011PhR...505....1K,2012PhR...517..141Y, Mack:2001gc} and have tried constraining parameters of such PMFs.

In this article, providing the PL as the distribution function of the PMF
(PL-PMF), we report effects of the PL-PMF on primordial
abundances of light elements up to Li as a function of the baryon to
photon ratio $\eta$.  The energy density of the PL-PMF is dependent on
the field strength, the PL index, and the maximum and minimum scales of the PMFs at the generation epochs.
Therefore, we also investigate degeneracies of these PL-PMF parameters in the effects of the PMF on light element abundances in detail for the first time.

In Sec.II, we describe how to include effects of the PL-PMF on BBN in a
code for the standard BBN (SBBN) model.
In Sec.III, we release the numerical computational results of BBN with the PL-PMF, and we discuss comparisons between our results and the observational constraints on elemental abundances in Sec.IV. In Sec.V, we summarize our study and mention future works.
\section{\label{sec:models}Model}
In this section, we describe our code for the BBN calculation with recent
updates, and suggest how to consider the effects of the PL-PMF on BBN. We only
consider the effect of boosts of cosmic expansion by the PMF energy
density because this effect is more important than other effects such as
changes in the rate of weak
\cite{1996PhRvD..54.4714C,1996PhLB..379...73G,1996PhRvD..54.7207K} and
nuclear \cite{2012arXiv1204.6164K} reactions, and those in the distribution
functions of electrons and positrons \cite{1996PhRvD..54.7207K}, which were
taken into account in Ref. \cite{2012arXiv1204.6164K}.
We modify the Kawano's BBN code
\cite{Kawano_Code1992,1993ApJS...85..219S}, taking into consideration the PMF
effects, and use the updated reaction rates with the recommendations of the JINA REACLIB Database V1.0 \cite{2010ApJS..189..240C} as adopted in Ref.
\cite{2012arXiv1204.6164K}. 
We use $\tau_\mathrm{n} = 878.5 (\pm
0.7_\mathrm{st} \pm 0.3 _\mathrm{sy})$ s as the neutron lifetime
\cite{2010PhRvC..82c5501S}, which is estimated with an improved
measurement \cite{2005PhLB..605...72S}.  This lifetime is somewhat
shorter than the average neutron lifetime of $880.1 \pm 1.1$~s
calculated by the Particle Data Group \cite{Beringer:1900zz}.  It better
satisfies the unitarity test of the Cabibbo-Kobayashi-Maskawa matrix
\cite{2005PhLB..605...72S}, and it slightly affects the primordial
abundances (most significantly that of
$^4$He)~\cite{2005PhRvD..71b1302M}.  Effects of the difference between
this lifetime and a longer one $885.7 \pm 0.8$~s from the old
recommendation by the Particle Data Group \cite{2010JPhG...37g5021N} are
seen in Figs. 1 and 2 in Ref. \cite{2012arXiv1204.6164K}.  The
uncertainties in the calculated abundances originating from an uncertainty
in the neutron lifetime are much smaller than both uncertainties in
observationally induced abundances and changes in calculated abundances
caused by the magnetic field.  We, therefore, do not take into account the 
uncertainty in the neutron lifetime.
\subsection{\label{subsec:models1}Constraints on primordial elemental abundances}
We shall describe constraints on the primordial elemental abundances that we adopt in this article. 
The primordial deuterium abundance is derived by observing the Lyman-$\alpha$ absorption systems of the quasi-stellar objects (QSOs).  A recent
measurement of the deuterium abundance in a damped Lyman alpha system of QSO
SDSS J1419+0829 was performed more precisely than those of any other QSO
absorption system~\cite{2012MNRAS.425.2477P}.  We then adopt two
constraints: (1) a mean value of ten QSO absorption systems including
J1419+0829, and (2) the abundance of J1419+0829 from the best measurement, as
follows,
\begin{eqnarray}
2.40 \times 10^{-5} < \frac{\mathrm{D}}{\mathrm{H}} < 2.88 \times 10^{-5}~
\nonumber \\
\mathrm{(2\sigma~{\rm range~from~mean~value})}, \label{limited_D_mean}
\\
2.43 \times 10^{-5} < \frac{\mathrm{D}}{\mathrm{H}} < 2.64 \times 10^{-5}~
\nonumber \\
\mathrm{(2\sigma~{\rm range~from~best~value})}.
\label{limited_D}
\end{eqnarray}
 The two groups IT10 \cite{2010ApJ...710L..67I} and AOS10 \cite{2010JCAP...05..003A} have reported different results for the primordial abundance of helium, $Y_\mathrm{p}$, from surveys of metal-poor extragalactic HII regions. They obtained the following constraints, i.e.,
\begin{eqnarray}
\mathrm{IT10:}~0.2463
 < Y_\mathrm{p}
 < 0.2667~\mathrm{(2\sigma)}
,\label{limited_Yp_IT}
\\
\mathrm{AOS10:}~0.2345  <
 Y_\mathrm{p}
 < 0.2777~\mathrm{(2\sigma)}
.
\label{limited_Yp_AOS}
\end{eqnarray}
Observations of the metal-poor halo stars (MPHSs) provide a constraint on the abundance of $^7$Li \cite{2010A&A...522A..26S}, i.e.,
\begin{eqnarray}
1.06 \times 10^{-10}
 < \frac{\mathrm{^7Li}}{\mathrm{H}}
 < 2.35 \times 10^{-10}~\mathrm{(2\sigma)}.
\label{limited_7Li}
\end{eqnarray}

The Wilkinson Microwave Anisotropy Probe (WMAP) project, which observes the CMB across the full sky, constrains the baryon density parameter \cite{2011ApJS..192...16L}, $\Omega_\mathrm{b}h^2= 0.02258 ^{+0.00114}_{-0.00112} $ ($2\sigma$). This result corresponds to the baryon to photon number ratio, $\eta/10^{-10} = 6.225^{+0.314}_{-0.309}$ ($2\sigma$).
\subsection{\label{sec:models}Cosmic Expansion Rate and BBN}
In this subsection, at first, we introduce the cosmic expansion. Next we describe an effect of the cosmic expansion on BBN. 
Tthe Hubble parameter $H$ for the homogeneous and isotropic flat-Universe is derived by the Friedmann equation and the conservation of energy momentum tensor,
\begin{eqnarray}
\left(
\frac{\dot{a}}{a}
\right)^2
 \equiv H^2 = \frac{8\pi G}{3}\rho \label{eq_friedmann},\\
\frac{d\rho}{dt} = -3H
\left(
	\rho + \frac{p}{c^2}
\right)
\label{eq_conservation},
\end{eqnarray}
where the dot expresses the time derivative as $\dot{a} = da/dt$, 
$G$ is the gravitational constant, $c$ is the photon speed, $\rho$ and $p$ are the energy density and the pressure, respectively, in the Universe.
The inverse of $H$ corresponds to the age of the Universe, $t_\mathrm{age} \sim H^{-1}$.
We assume that the main components of the Universe are photon, neutrino, baryon and cold dark matter (CDM), they are the ideal fluid, and the Universe expands adiabatically.  

We describe a rate of reaction between some particles as $\Gamma$.  The time scale  of the reaction is then $\Gamma^{-1}$. When the reaction time scale becomes larger than the age of the Universe, i.e., $t_\mathrm{age} \sim H^{-1}$, the reaction effectively stops in the Universe.  Whether a reaction is operative or not is determined by the following condition, i.e.,
\begin{eqnarray}
\Gamma^{-1} < H^{-1},~\Gamma > H&:& \mathrm{~the~reaction~continues}, \label{Gamma_H1}\\
\Gamma^{-1} > H^{-1},~\Gamma < H&:& \mathrm{~the~reaction~stops}.\label{Gamma_H2}
\end{eqnarray}
 Since the final nuclear abundances of BBN depend on when the respective reactions become ineffective, the energy densities of the Universe influence the abundances of the elements produced in BBN.
\subsection{\label{subsec:models_ED_PLPMF}Energy density of power law PMF}
We will explain how the PL-PMF energy density affects the cosmic expansion and BBN.
In the early Universe including the BBN epoch, we can assume that statistically average motions of fluids are negligibly small. 
In this case, from Maxwell's equations and Ohm's law, we derive the induction equation \cite{Dendy:1990booka}
\begin{eqnarray}
\frac{\partial \mbi{B}}{\partial t}=\zeta\nabla^2 \mbi{B}.\label{eq_MO}
\end{eqnarray}
Here $\zeta$ is the magnetic diffusivity, which is defined by
\begin{eqnarray}
\zeta \equiv \frac{c^2}{4\pi \sigma} = \frac{c^2\omega_e}{4\pi}
\end{eqnarray}
where $\sigma$ is the electric conductivity and $\omega_e$ is the electrical resistance \cite{Dendy:1990booka} as follow,
\begin{eqnarray}
\omega_e
\equiv
\frac{1}{\sigma}
     =
  \frac{m_e}{n_e e^2}
   c n_\gamma\sigma_\mathrm{T}
=
  \frac{c \sigma_\mathrm{T} m_e }{\eta e^2}
 \label{sigma}
\end{eqnarray}
Here $m_e$ is the mass of the electron, 
$e$ is the electron charge,
$\sigma_\mathrm{T}$ is the Thomson scattering cross section, 
$n_\gamma$ is the number density of the photon, $n_e$ is the number density
of the electron, and $\eta$ is the baryon to photon ratio.
Equation~(\ref{eq_MO}) describes the dissipation of the magnetic field, and it shows that the magnetic field dissipates rapidly with time on the scale $L < \sqrt{t_\mathrm{age}(T)\zeta}$ \cite{Grasso:2000wj,Dendy:1990booka}, 
where $T$ is the temperature, and $t_\mathrm{age}$ is the age of the Universe which is defined by $t_\mathrm{age} \sim 1.32 \times (1\mathrm{MeV}/T)^2$ s at the relativistic degrees of freedom, $g_* = 3.36$.
Therefore, the magnetic field is hard to be generated on the scale $L \ll \sqrt{t_\mathrm{age}(T)\zeta} = L_\mathrm{FI}(T)$.
On the other hand, the magnetic field on a scale much larger than $L_\mathrm{FI}(T)$ is hard to dissipate in the Universe by time $t_\mathrm{age}$, and the magnetic field is "frozen-in" the dominant fluids \cite{Dendy:1990booka}.
In this article, we are interested in magnetic fields which survive until the end of BBN.
In the epoch during which the electron positron annihilation has been completed ($T \sim 0.01$MeV), the cutoff scale for PMF generations, $L_\mathrm{FI}(T)$ for $10^{-10}<\eta<10^{-9}$, is 
\begin{eqnarray}
2.73\times 10^{5}
\mathrm{cm}
\times
\left(
  \frac{0.01 \mathrm{MeV}}{T}
\right)^{\frac{3}{4}}
\le L_\mathrm{FI}(T)
\nonumber \\
\le
8.63\times 10^{5}
\mathrm{cm}
\times
\left(
  \frac{0.01 \mathrm{MeV}}{T}
\right)^{\frac{3}{4}},
\end{eqnarray}
the comoving scale $L^0_\mathrm{FI}~=~ L_\mathrm{FI}(T=0.01\mathrm{MeV}) / a(T=0.01\mathrm{MeV})$ is 
\begin{eqnarray}
1.16\times 10^{13}
\mathrm{cm}
\le L^0_\mathrm{FI} \le
3.67\times 10^{13}
\mathrm{cm}
\end{eqnarray}
and the comoving wave number $k_\mathrm{FI} = 2\pi/L^0_\mathrm{FI}$ is
\begin{eqnarray}
1.72
\times 10^{-12}
\mathrm{cm}^{-1}
\le k_\mathrm{FI} \le
5.41
\times 10^{-13}
\mathrm{cm}^{-1}.
\end{eqnarray}
Here we use $a(T=0.01\mathrm{MeV}) = T_\mathrm{CMB}/0.01\mathrm{MeV} \sim 2.35 \times 10^{-8}$.
Thus we will report results of the effects on BBN from the PMF with a comoving wave number less than $k_\mathrm{FI} \sim 10^{-13}~\mathrm{cm}^{-1}$.

The PL spectrum is one of the most familiar forms for distributions of the energy densities including that of the PMF \cite{Grasso:2000wj,Mack:2001gc,2011PhR...505....1K,2012PhR...517..141Y}.  It appears in various physical phenomena operating on the cosmological scale. Many researchers have been studying the effect of the PMF with the PL spectrum on various physical processes in the early Universe \cite{Grasso:2000wj,Mack:2001gc,2011PhR...505....1K,2012PhR...517..141Y}, and have tried to constrain parameters of such PMFs. 
We then assume that the PMFs with the PL spectrum are statistically homogeneous, isotropic, and random, as assumed in previous studies, and we research their effects on BBN.

After a magnetic field is generated, the total energy density originating from that of the magnetic field is conserved, even if part of the magnetic field energy is transferred to other physical degrees of freedom (e.g. kinetic or/and thermal energies of dominant fluids or the gravitational wave) in the adiabatic Universe. 
The cosmic expansion depends on the total energy density in the Universe. Therefore, when we consider the effect of the PMF energy density on the cosmic expansion, it is reasonable to define the total energy density originating from the PMF at its generational epoch.
We shall discuss cases of finite energy transfer from the PMF to other physical degrees of freedom in Sec.~\ref{sec:discussions}.

Since the strength of the frozen-in PMF is proportional to an inverse square of the scale factor, its energy density and pressure depend on the inverse fourth power of the scale factor as follows \cite{2007PhR...449..131B,Yamazaki:2006mi,2011PhR...505....1K,Yamazaki:2007oc,2010AdAst2010E..80Y,2012PhR...517..141Y}:
\begin{eqnarray}
\rho_\mathrm{PMF} = \frac{3}{c^2}p_\mathrm{PMF} \propto \frac{B^2(a_0)}{a^4}.
\end{eqnarray}
We can neglect the pressure and energy density of matter in the
radiation dominated era.
Therefore, the total energy density and pressure in the early Universe are
\begin{eqnarray}
\rho \sim \rho_\mathrm{R}+ \rho_\mathrm{PMF}, \label{eq_rhos}\\
p \sim p_\mathrm{R} + p_\mathrm{PMF}.\label{eq_Ps}
\end{eqnarray}

Next, we shall derive the energy density of the PL-PMF.
A two-point correlation function of the strength of the PMF \cite{Mack:2001gc,Yamazaki:2006mi,2011PhR...505....1K,Yamazaki:2007oc,2010AdAst2010E..80Y,2012PhR...517..141Y} is given by
\begin{eqnarray}
\left\langle 
	B^{i}(\mbi{k}) {B_{j}}^*(\mbi{k}')
\right\rangle 
	&=& 
	(2\pi)^3P_\mathrm{[PMF]}(k)P^{i}_j(k)\delta(\mbi{k}-\mbi{k}')~.
\label{eq:two_point_correlations_of_PMF}
\end{eqnarray}
Here
\begin{eqnarray}
P^{i}_j(k)&=&
	\delta^{i}_j-\frac{k{}^{i}k{}_{j}}{k{}^2} \label{project_tensor}
\end{eqnarray}
and
\begin{eqnarray}
P_\mathrm{[PMF]}(k) = Ak^{n_\mathrm{B}}. \label{eq_power1}
\end{eqnarray}
The convention for the Fourier transform is
\begin{eqnarray}
f(\mbi{k}) = \int \mathrm{exp} (i\mbi{k} \cdot \mbi{x}) F(\mbi{x}) d^3 x .
\end{eqnarray}
In this article we need the energy density of the PMF in the Universe, $\rho_\mathrm{PMF} = \langle B^2 \rangle /(8\pi)$. 
From Eq.~(\ref{eq:two_point_correlations_of_PMF}), we obtain
\begin{eqnarray}
\left\langle 
	B^{i}(\mbi{k}) {B_{i}}^*(\mbi{k}')
\right\rangle 
&=& 2(2\pi)^3P_\mathrm{[PMF]}(k)\delta(\mbi{k}-\mbi{k}')
\nonumber \\
&=& 2(2\pi)^3 A k^{n_\mathrm{B}}\delta(\mbi{k}-\mbi{k}').
\label{eq:two_point_correlations_of_PMF2}
\end{eqnarray}
We derive the coefficient from the variance of the magnetic fields $A$ in real space.  First we define
\begin{eqnarray}
\left.
	\left\langle
		B^{i}(\mbi{x}) {B_i}(\mbi{x})
	\right\rangle
\right|_\lambda
= B_\lambda^2.
\label{eq_real}
\end{eqnarray}
Here $\lambda$ is a comoving radius for a Gaussian sphere, and $B_\lambda$ is a comoving PMF strength which is scaled to the present-day value on some length scale $\lambda$. 
From Eqs.~(\ref{eq:two_point_correlations_of_PMF})~-~(\ref{eq_power1})~and~(\ref{eq_real}),
\begin{eqnarray}
\left.
	\left\langle
		B^{i}(\mbi{x}) {B_i}(\mbi{x})
	\right\rangle
\right|_\lambda
&=&
B_\lambda^2
\nonumber \\
&=&
	\frac{1}{(2\pi)^6}
	\int d^3 k 
	\int d^3 k'
		\exp(-i\mbi{x}\cdot\mbi{k}+i\mbi{x}\cdot\mbi{k}') 
\nonumber \\
&&\times 
		\left\langle
			B^{i}(\mbi{k}) {B_i^{\ast}}(\mbi{k}')
		\right\rangle
\times |W_{\lambda}^2(k)|,\label{eq_real2}
\end{eqnarray}
where $W_\lambda(k)$ is a window function here assumed to be a Gaussian filter, 
$W_\lambda(k) = \exp(-\lambda^2k^2/2)$. We then obtain the coefficient $A$ as follows
\begin{eqnarray}
A &=& 
B^2_\lambda
\frac{(2\pi)^2}{4}
\left(
   \int dk k^{n_\mathrm{B}+2} \exp(-\lambda^2k^2)
\right)^{-1}
\nonumber \\
&=&
B^2_\lambda
\frac{(2\pi)^2}{2}
\frac{
\lambda^{n_\mathrm{B}+3}
}
{
   \Gamma
   \left(
      \frac{n_\mathrm{B}+3}{2}
   \right)
}
\nonumber \\
&=&
B^2_\lambda
\frac{(2\pi)^{n_\mathrm{B}+5}}{2}
\frac{1}
{
   k_\mathrm{[PMF]}^{n_\mathrm{B}+3}
   \Gamma
   \left(
      \frac{n_\mathrm{B}+3}{2}
   \right)
},
\label{eq_A}
\end{eqnarray}
where $\lambda=2\pi/k_\mathrm{[PMF]}$.
Therefore
\begin{eqnarray}
P_\mathrm{[PMF]}(k)
= 
\frac{
(2\pi)^2 B^2_\lambda
\lambda^{n_\mathrm{B}+3}
}
{
   2
   \Gamma
   \left(
      \frac{n_\mathrm{B}+3}{2}
   \right)
}
k^{n_\mathrm{B}}
\label{eq:two_point_correlations_of_PMF3}
\end{eqnarray}
From Eqs~(\ref{eq:two_point_correlations_of_PMF})~-~(\ref{eq:two_point_correlations_of_PMF2})~and~(\ref{eq:two_point_correlations_of_PMF3}),
the energy density of the PMF is
\begin{eqnarray}
\rho_B
 &=& 
\frac{\langle B^2 \rangle}{8\pi}
 = 
\frac{2}{8\pi}
\int^{k_\mathrm{[max]}}_{k_\mathrm{[min]}}
\frac{dk}{k}
\frac{k^3}{2\pi^2}
P_\mathrm{[PMF]}(k)
\nonumber \\ 
 &=& 
\frac{2}{8\pi}
\int^{k_\mathrm{[max]}}_{k_\mathrm{[min]}}
\frac{dk}{k}
\frac{k^3}{2\pi^2}
\frac{
(2\pi)^2 B^2_\lambda
\lambda^{n_\mathrm{B}+3}
}
{  2
   \Gamma
   \left(
      \frac{n_\mathrm{B}+3}{2}
   \right)
}
k^{n_\mathrm{B}}
\nonumber \\ 
 &=& 
\frac{1}{8\pi}
\frac{
B^2_\lambda
}
{
   \Gamma
   \left(
      \frac{n_\mathrm{B}+5}{2}
   \right)
}
\left[
(\lambda k_\mathrm{[max]})^{n_\mathrm{B}+3}
-
(\lambda k_\mathrm{[min]})^{n_\mathrm{B}+3}
\right], 
\nonumber \\
\label{eq:PL_PMF_energy_density}
\end{eqnarray}
where $k_\mathrm{[max]}$ and $k_\mathrm{[min]}$ are the maximum and
minimum wave numbers, respectively, which depend on generation models.

The first purpose of this study is to investigate the effects of the PMF energy density on BBN and to constrain the PMF energy density in the BBN epoch. 
The second purpose is to discuss of the degeneracy between the PL-PMF parameters and distribution models of the PMF.
To effectively proceed with these studies and discussions, we redefine the scale invariant (SI) field strength of the PMF from Eq.~(\ref{eq:PL_PMF_energy_density}) , which is constrained from observational constraints on elemental abundances, i.e.,
\begin{eqnarray}
B^\mathrm{X}_\mathrm{SI}
\equiv
B^\mathrm{X}_\lambda
\sqrt{
\frac{
\left[
(\lambda k_\mathrm{[max]})^{n_\mathrm{B}+3}
-
(\lambda k_\mathrm{[min]})^{n_\mathrm{B}+3}
\right]
}
{
   \Gamma
   \left(
      \frac{n_\mathrm{B}+5}{2}
   \right)
}
}
, 
\end{eqnarray}
and 
\begin{eqnarray}
B^\mathrm{X}_\lambda (n_\mathrm{B}, k_\mathrm{[max]}, k_\mathrm{[min]}) = 
B^\mathrm{X}_\mathrm{SI}
   \sqrt{
   \frac{
   \Gamma \left(\frac{n_\mathrm{B}+5}{2}\right)
   }
   {
   \left(
      k_\mathrm{[max]} ^{n_\mathrm{B}+3}
     -
      k_\mathrm{[min]} ^{n_\mathrm{B}+3}
   \right)
   \lambda^{n_\mathrm{B}+3}
}
},
\nonumber \\
\label{eq:field_strength_for_elements}
\end{eqnarray}
where X is a species of light element produced in BBN and 
$B^\mathrm{X}_\mathrm{SI}$ represents the corresponding value of the SI PMF strength. Since $B^\mathrm{X}_\mathrm{SI}$ is proportional to 
$\sqrt{\rho_\mathrm{PMF}}$ and does not depend on other PMF parameters, it is useful to directly understand effects of the PL-PMF amplitude on BBN. Furthermore it is relatively easy to discuss the degeneracy of the PMF parameters if we use this formalization.
\section{\label{sec:resultss}results}
In this section, we show calculated results of the BBN with the PL-PMF using our code.
We also show how the strength of the PL-PMF is constrained from limits on abundances of light elements up to Li.
\subsection{\label{subsec:BBNwithPMF} BBN with PL-PMF}
Figure~\ref{fig1} shows abundances of $^4$He ($Y_{\rm p}$: mass fraction), D/H, $^3$He/H, $^7$Li/H and $^6$Li/H as a function of $\eta$ for cases of BBN with (solid lines) and without (dashed lines) the effects of the PL-PMF.  The parameters of the PL-PMF are set to be $B_\mathrm{1~Mpc}=2.2~\mu$G, $n_{\rm B}=-2.99$, $k_{\rm [max]}=10^{-13}$ cm$^{-1}$, and $k_{\rm [min]}=0$ cm$^{-1}$.  This parameter set corresponds to a SI PMF strength of $B^\mathrm{X}_\mathrm{SI}=2.54~\mu$G.
The vertical painted band shows the limit on the baryon to photon ratio derived from WMAP 7yr data \cite{2011ApJS..192...16L}. 
The horizontal painted bands and lines with downward arrows indicate observational constraints and upper limits, respectively, on elemental abundances.  

Effects of the PMF on abundances of respective nuclides have been analyzed in detail \cite{2012arXiv1204.6164K}.  The effects are described below to explain the abundances predicted as a function of $\eta$ in Fig.~\ref{fig1}.

When the PL-PMF exists, abundances of the neutron are larger than those of the BBN without the PL-PMF, i.e., the SBBN. The reason is as follows: Considering the energy density of the PL-PMF, the cosmic expansion rate $H$ is larger than that of SBBN.  Therefore, the weak reaction freeze-out occurs earlier \cite{1999PhRvD..59l3002S} [cf. Eqs.~(\ref{Gamma_H1})~and~(\ref{Gamma_H2})], and abundances of the neutron at the freeze-out increase. Furthermore, the faster expansion leads to a shorter time interval between the freeze-out and $^4$He production, so abundances of the neutron, which is spontaneously decaying after the freeze-out, are higher at the epoch of $^4$He production.  Since $^4$He nuclei are formed from almost all neutrons, the energy density of the PL-PMF increases the abundance of $^4$He [Fig.~\ref{fig1}(a)].

Deuterium ($^2$H, D) is produced and destroyed by
\begin{eqnarray}
 p + n &\rightarrow& ^2{\rm H} + \gamma, \label{pn_Dg}\\
^2{\rm H} + ^2{\rm H} &\rightarrow& ^3{\rm H} + p,\label{DD_Tp} \\
^2{\rm H} + ^2{\rm H} &\rightarrow& ^3{\rm He} + n.\label{DD_3Hen}
 \end{eqnarray}
The high neutron abundance in the PL-PMF case leads to an increase in D abundance because of an enhanced production rate through the reaction (\ref{pn_Dg}) [Fig.~\ref{fig1}(b)].

$^3$H and $^3$He are produced via the reactions (\ref{DD_Tp}) and (\ref{DD_3Hen}), and destroyed by
\begin{eqnarray}
^3{\rm H} + ^2{\rm H} &\rightarrow& ^4{\rm He} + n, \label{TD_4Hen} \\
^3{\rm He} + n &\rightarrow& ^3{\rm H} + p. \label{3Hen_3Hp}
\end{eqnarray}
The enhanced abundances of the neutron and D result in higher $^3$H and slightly higher $^3$He abundances than those of SBBN.  $^3$H nuclei eventually $\beta^-$ decay to $^3$He.  The final primordial abundance of $^3$He is then the sum of those of $^3$H and $^3$He in the BBN epoch. Primordial $^3$He is mainly produced as $^3$He (in SBBN more than 99.5 \% of the final abundances of $^3$He result from those of $^3$He for $\eta=10^{-9}$--$10^{-10}$).  The slight increase in $^3$He at the BBN epoch enhances the final $^3$He abundance slightly [Fig.~\ref{fig1}(b)].

$^7$Li is mainly produced and destroyed by
\begin{eqnarray}
^4{\rm He} + ^3{\rm H} &\rightarrow& ^7{\rm Li} + \gamma, \label{4HeT_7Ligamma}\\
^7{\rm Li} + p &\rightarrow& ^4{\rm He} + ^4{\rm He}.\label{7Lip_4He4He}
\end{eqnarray}
A combination of a significant increase in $^3$H abundance and a small fractional increase in $^4$He abundance, both by the PL-PMF, results in a significant increase in the $^7$Li production rate [Eq.~(\ref{4HeT_7Ligamma})], while a small fractional decrease in H abundance (its mass fraction is $X_{\rm p}\approx 1-Y_{\rm p}$) results in a slight decrease in the $^7$Li destruction rate [Eq.~(\ref{7Lip_4He4He})].  As a result, $^7$Li abundance is larger than that in SBBN.

$^7$Be is mainly produced and destroyed by
\begin{eqnarray}
^4{\rm He} + ^3{\rm He} &\rightarrow& ^7{\rm Be} +\gamma, \label{4He3He_7Begamma} \\
^7{\rm Be} + n &\rightarrow& ^7{\rm Li} + p.\label{7Ben_7Lip}
\end{eqnarray}
A significant increase in neutron abundance by the PL-PMF results in an increase in the $^7$Be destruction rate [Eq.~(\ref{7Ben_7Lip})], while a combination of a slight increase in $^3$He abundance and a small fractional increase in $^4$He abundance results in a slight increase in the $^7$Be production rate [Eq.~(\ref{4He3He_7Begamma})].  As a result, $^7$Be abundance is smaller than that in SBBN.

$^7$Be nuclei eventually capture electrons and are converted to $^7$Li nuclei at the epoch of cosmological recombination.  The primordial $^7$Li abundance is, therefore, contributed from both $^7$Li and $^7$Be produced at BBN.   If the baryon to photon ratio $\eta$ is lower than some critical value $\eta_{\rm crit}$, abundance of $^7$Li is larger than that of $^7$Be.  On the other hand, if $\eta$ is larger than $\eta_{\rm crit}$, abundance of $^7$Be is larger than that of $^7$Li.  Therefore, the equality, i.e., $^7$Li/H=$^7$Be/H, is satisfied at $\eta=\eta_{\rm crit}$.  Since the PMF increases $^7$Li abundance, while decreases $^7$Be abundance, the final $^7$Li abundances are higher in $\eta \lesssim \eta_{\rm crit}$ and lower in $\eta \gtrsim \eta_{\rm crit}$ than those in SBBN.  It should be noted that the $\eta_{\rm crit}$ value and the minimum value of the final $^7$Li abundance, which is located at a valley of the curve for $^7$Li/H, are moved from those of SBBN by the PMF effects:  The 
 values of SBBN are $\eta_{\rm crit}= 3.00 \times10^{-10}$ and $^7$Li/H$_{\rm min}=1.18\times10^{-10}$ at $\eta=2.50 \times 10^{-10}$, while those of BBN with the PL-PMF are $\eta_{\rm crit}= 3.71 \times10^{-10}$ and $^7$Li/H$_{\rm min}=1.41\times10^{-10}$ at $\eta=3.10 \times 10^{-10}$ [Fig.~\ref{fig1}(c)].

$^6$Li is produced and destroyed by
\begin{eqnarray}
^4{\rm He} + ^2{\rm H} &\rightarrow& ^6{\rm Li} +  \gamma,\label{4HeD_6Ligamma} \\
^6{\rm Li} + p &\rightarrow& ^3{\rm He} + ^4{\rm He}.\label{6Lip_3He4He}
\end{eqnarray}
The relatively large enhancement of D abundance and a small fractional increase in $^4$He result in an efficient production of $^6$Li [Fig.~\ref{fig1}(d)].

The $^4$He abundance increases as a function of $\eta$, and it increases
as a function of the PMF amplitude. The limits on $^4$He abundance,
therefore, prefer an $\eta$ value smaller than in the SBBN when the
effect of the PMF is considered.  The D abundance is smaller for larger
$\eta$ values, while it is larger for larger amplitudes of the PMF.  The
limits on D abundance then prefer an $\eta$ value larger than in the SBBN.  The $^7$Li abundance is larger for larger $\eta$ values, while it
is smaller for larger amplitudes of the PMF.  The limit on $^7$Li
abundance, therefore, prefers an $\eta$ value larger than in the SBBN when the effect of the PMF is considered (see Sec.~\ref{subsec:discusses2} also).
\subsection{\label{subsec:discussions2} Constraint on the PL-PMF from light element abundances}
To research not only qualitative results but also the quantitative relationship between $B^\mathrm{X}_\mathrm{SI}$ and $\eta$, we show how the PL-PMF strength $B^\mathrm{X}_\mathrm{SI}$ can be constrained via observational constraints on the abundance of light elements [Eqs.~(\ref{limited_D_mean})~(\ref{limited_D}),~(\ref{limited_Yp_IT}),~(\ref{limited_Yp_AOS})~and~(\ref{limited_7Li})]
and the $\eta$ value.
Figure \ref{fig2} shows calculated constraints of $B^\mathrm{X}_\mathrm{SI}$s as a function of $\eta$. The vertical painted band is the limit on the baryon to photon ratio from the WMAP 7yr data \cite{2011ApJS..192...16L}.  
We use abundances of only $^4$He, D, and $^7$Li to constrain the $B^\mathrm{X}_\mathrm{SI}$ value since limits on $B^\mathrm{X}_\mathrm{SI}$ from abundances of $^3$He and $^6$Li are less severe than those of $^4$He, D, and $^7$Li [Figs.~\ref{fig1}(b) and (d)]. 
Four kinds of lines in Fig.~\ref{fig2}(a) correspond to upper limits, i.e., $B^\mathrm{X}_\mathrm{SI}$, 
from 
$Y_\mathrm{p}$ $<$ 0.2777 (AOS10; black-thin-dotted-dashed line), 
$Y_\mathrm{p}$ $<$ 0.2667 (IT10; green-thick-dotted-dashed line), 
D/H $<$ 2.88 $\times 10^{-5}$ (mean; blue-thin line), 
D/H $<$ 2.64 $\times 10^{-5}$ (best; red-thick line), and
 $^7$Li/H $<$ 2.35$\times 10^{-10}$ (purple-dotted line), respectively.
On the other hand, requiring a consistency between $^7$Li abundances of theory and observations leads to a lower limit on the PL-PMF as a function of $\eta$ [the right purple-dotted line in Fig.~\ref{fig2}(a)] because the abundance of $^7$Li decreases when the energy density of the PL-PMF increases (Sec.\ref{subsec:BBNwithPMF}).
Then, the allowed region from the $^7$Li limit on the $B^\mathrm{^7Li}_\mathrm{SI}$ vs. $\eta$ plane is between the two purple-dotted curves in Fig.~\ref{fig2}(a).

The limits on PMF strengths, i.e, $B^\mathrm{X}_\mathrm{SI}$,
at the $\eta$ value of a 2 $\sigma$ lower limit by the WMAP 7yr data \cite{2011ApJS..192...16L}, i.e, $\eta$=5.916 $\times 10^{-10}$, are
\begin{eqnarray}
B^{Y_\mathrm{p}}_\mathrm{SI}
&\le&
2.00~\mu\mathrm{G~~for~}Y_\mathrm{p}\mathrm{\le0.2667~(IT10)}, 
\label{lower_B_Yp_IT}
\\
B^{Y_\mathrm{p}}_\mathrm{SI}
&\le&
2.56~\mu\mathrm{G~~for~}Y_\mathrm{p}\mathrm{\le0.2777~(AOS10)}, 
\label{lower_B_Yp_AOS}
\\
B^\mathrm{D}_\mathrm{SI}
&\le&
0.848~\mu\mathrm{G~~for~D/H
\le2.88\times 10^{-5}~(mean)}, 
\label{lower_B_D}
\\
B^\mathrm{^7Li}_\mathrm{SI}
&\ge&
5.05~\mu\mathrm{G~~for~ ^7Li/H\le 2.35\times 10^{-10}}, 
\label{lower_B_7Li}
\\
&&\mathrm{at~} \eta = 5.916 \times 10^{-10}. \nonumber
\end{eqnarray}
Since the abundance of D/H is already more than D/H = $2.64\times 10^{-5}$ (best) when no PMF is assumed, we cannot put a constraint ($B^\mathrm{D}_\mathrm{SI}$) using
D/H
$\le2.64\times 10^{-5}$ at 
$\eta = 5.916 \times 10^{-10}$, 

When we adopt the $\eta$ values of the best-fit and 2 $\sigma$ upper limit by the WMAP 7yr data \cite{2011ApJS..192...16L}, constraints from $^7$Li abundance are weaker, or $B^\mathrm{^7Li}_\mathrm{SI}$ values are larger. 
We then show four constraints ($B^\mathrm{X}_\mathrm{SI}$), except for that of $^7$Li as follows:
\begin{eqnarray}
B^{Y_\mathrm{p}}_\mathrm{SI}
&\le&
1.97~\mu\mathrm{G~~for~}Y_\mathrm{p}\mathrm{\le0.2667~(IT10)}, 
\label{best_B_Yp_IT}
\\
B^{Y_\mathrm{p}}_\mathrm{SI}
&\le&
2.54~\mu\mathrm{G~~for~}Y_\mathrm{p}\mathrm{\le0.2777~(AOS10)}, 
\label{best_B_Yp_AOS}
\\
B^\mathrm{D}_\mathrm{SI}
&\le&
1.51~\mu\mathrm{G~~for~D/H
\le2.88\times 10^{-5}~(mean)}, 
\label{best_B_D1}
\\
B^\mathrm{D}_\mathrm{SI}
&\le&
0.780~\mu\mathrm{G~~for~D/H
\le2.64\times 10^{-5}~(best)}, 
\label{best_B_D2}
\\
&&\mathrm{at~} \eta = 6.225 \times 10^{-10}, \nonumber
\end{eqnarray}
and 
\begin{eqnarray}
B^{Y_\mathrm{p}}_\mathrm{SI}
&\le&
1.95~\mu\mathrm{G~~for~}Y_\mathrm{p}\mathrm{\le0.2667~(IT10)}, 
\label{upper_B_Yp_IT}
\\
B^{Y_\mathrm{p}}_\mathrm{SI}
&\le&
2.51~\mu\mathrm{G~~for~}Y_\mathrm{p}\mathrm{\le0.2777~(AOS10)}, 
\label{upper_B_Yp_AOS}
\\
B^\mathrm{D}_\mathrm{SI}
&\le&
1.97~\mu\mathrm{G~~for~D/H
\le2.88\times 10^{-5}~(mean)}, 
\label{upper_B_D1}
\\
B^\mathrm{D}_\mathrm{SI}
&\le&
1.45~\mu\mathrm{G~~for~D/H
\le2.64\times 10^{-5}~(best)}, 
\label{upper_B_D2}
\\
&&\mathrm{at~} \eta = 6.539 \times 10^{-10}. \nonumber
\end{eqnarray}
This result is very similar to the latest result \cite{2012arXiv1204.6164K} of the BBN calculation which consistently includes effects of the PMF through changes in the distribution functions of electrons and positrons as well as the cosmic expansion rate.  This is because the effect of the PMF on BBN through changes in distribution functions is smaller than that through the cosmic expansion rate \cite{1996PhRvD..54.7207K}.

Adopting our results [(\ref{lower_B_Yp_IT})-(\ref{upper_B_D2}) and Fig.~\ref{fig2}]
and the WMAP-$\eta$ value \cite{2011ApJS..192...16L}, 
we show combined constraints on the PL-PMF [Fig.~\ref{fig2}(b)].
If we adopt limits on abundances of $Y_\mathrm{p}$(IT10) and D/H (mean), 
we obtain  a combined upper limit:
\begin{eqnarray}
B^{Y_\mathrm{p} + \mathrm{D}}_\mathrm{SI}
&\le&
1.95 
~\mu\mathrm{G}
\nonumber \\
&\mathrm{for}&
Y_\mathrm{p}~\mathrm{\le 0.2667~(IT10)}
\nonumber \\
&&
\mathrm{and~D/H~ \le 2.88\times10^{-5}~(mean)}.
\label{upper_B_Yp_IT1}
\end{eqnarray}
The maximum value corresponds to the limit at 
$\eta=6.523\times 10^{-10}$ (see the intersection of the thin solid curve and the thick dot-dashed curve in Fig.~\ref{fig2}).
If we adopt $Y_\mathrm{p}$ (IT10) and D/H (best), the following upper limit is derived:
\begin{eqnarray}
B^{Y_\mathrm{p} + \mathrm{D}}_\mathrm{SI}
&\le&
1.45 
~\mu\mathrm{G}
\nonumber \\
&\mathrm{for}&
Y_\mathrm{p}~\mathrm{\le 0.2667~(IT10)}
\nonumber \\
&&
\mathrm{and~D/H~\le 2.64\times10^{-5}~(best)}.
\label{upper_B_Yp_IT2}
\end{eqnarray}
The maximum value corresponds to the limit at $\eta=6.539\times
10^{-10}$ (see the intersection of the thick solid curve and the vertical painted band in Fig.~\ref{fig2}).
\section{\label{sec:discussions} Discussions}
In the above sections, we explained the effect of PL-PMF on BBN and  derived constraints on PMF energy densities.
In this section, considering these results, we will discuss degeneracy between the PL-PMF parameters, 
($B_\lambda,~ n_\mathrm{B},~ k_\mathrm{[max]},~ k_\mathrm{[min]}$)  in effects on BBN, how such constrained PL-PMF parameters are applied for constraint on PMF generation models, and cases of an energy transferring from PMF to others physical degrees of freedom.  
\subsection{\label{subsec:discussions1} Degeneracy between PL-PMF parameters}
The PL-PMF strength $B^\mathrm{X}_\lambda$ depends on $k_\mathrm{[max]}$, $k_\mathrm{[min]}$ and $n_\mathrm{B}$ [Eq.~(\ref{eq:field_strength_for_elements})].
In particular, if $k_\mathrm{[min]}/ k_\mathrm{[max]}$ is close to 1 or/and $n_B$  is close to -3 \cite{f_note_1},
the influence of $k_\mathrm{[min]}$ on $B^\mathrm{X}_\lambda$ cannot be ignored and 
we should treat the $k_\mathrm{[min]}$ value carefully in constraining PMF parameters and generation models (Fig.~\ref{fig3}), 
although $k_\mathrm{[min]}$ does not affect the energy density of the PL-PMF otherwise. 
If $k_\mathrm{[min]}/k_\mathrm{[max]}$ is close to 1, such a PMF is generated on narrow scale ranges. In this case,  the distribution of the PMF may be better described by the Gaussian, Delta or log-normal \cite{2011PhRvD..84l3006Y} functions than the PL.
When $n_\mathrm{B}$ is close to -3, the PMF is expected to be generated in the epoch of inflation \cite{2011PhR...505....1K,2010AdAst2010E..80Y,2012PhR...517..141Y}. 
In this case, from Eq.~(\ref{eq:field_strength_for_elements}) and Fig.~\ref{fig3} (b), the term of $k_\mathrm{[min]}^{n_\mathrm{B}+3}$ cannot be neglected even if $k_\mathrm{min}/ k_\mathrm{max}$ is relatively small.
Thus we need to be careful in constraining the PMF of inflationary origin.

We have discussed the specific cases of the PL-PMF.
Next, we discuss the opposite cases in which ranges of wave number for the generated PL-PMF spectrum are wide enough so that $k_\mathrm{[min]}$ can be neglected ($k_\mathrm{[min]}/k_\mathrm{[max]}\ll1$).
Figure \ref{fig4} shows constraints on the field strength $B_\lambda$ as a function of $n_\mathrm{B}$ [cf. Eqs.~(\ref{eq:field_strength_for_elements}) and (\ref{upper_B_Yp_IT1})] for fixed values of $\lambda = 1$~Mpc and $k_\mathrm{[min]} =  0$.
It is clear that the limits on this plane of $B_\lambda$ vs. $n_\mathrm{B}$ from light element abundances, i.e., their negative slopes, show a negative correlation of the parameters [Eq.~(\ref{eq:field_strength_for_elements})] in a parameter region for the fixed SI strength of the PMF.
When generated PL-PMFs have spectra with smaller $k_\mathrm{[max]}$, constraints on the ($B_\lambda$, $n_\mathrm{B}$) plane are more stringent.
In general, considering the cosmic expansion after the inflation, the smaller the generation scale of the magnetic field, the earlier the magnetic field is generated.
This feature is useful in constraining models for PMF generation in the post-inflation epoch.

Since cosmological observations have constrained strengths of the PL-PMF $B_\lambda$ at $\lambda=1$~Mpc that are less than orders of 1 nG \cite{Yamazaki:2004vq,Yamazaki:2006bq,Yamazaki:2008bb,Yamazaki:2006mi,2012PhR...517..141Y}, we shall discuss the degeneracy between $k_\mathrm{[max]}$ and $n_\mathrm{B}$ in the PMF effect on BBN for 0.1 nG  $< B_\lambda <$ 10 nG.
Figure \ref{fig5} shows constraints on the maximum wave number $k_\mathrm{[max]}$ as a function of $n_\mathrm{B}$ [cf. Eqs.~(\ref{eq:field_strength_for_elements}) and (\ref{upper_B_Yp_IT1})] for fixed values of $\lambda = 1$~Mpc and $k_\mathrm{[min]} =  0$.  Results for various $B_\lambda$ values of 10 nG (dashed line), 1 nG (solid line) and 0.1 nG (dot-dashed line) are shown.
It is obvious that the limits on the plane of $k_\mathrm{[max]}$ and $n_\mathrm{B}$ show a negative correlation of the parameters [Eq.~(\ref{eq:field_strength_for_elements})] in a parameter region for the fixed SI strength of the PMF.
The cosmological observations (e.g. the CMB and the large scale structures: see Refs. \cite{Yamazaki:2004vq,Yamazaki:2006bq,Yamazaki:2008bb,Yamazaki:2006mi,2012PhR...517..141Y} for details) have constrained the PMF with strength of the order of 1 nG at 1 Mpc.
As mentioned in Sec.~\ref{subsec:models_ED_PLPMF}, we are interested in the magnetic field on scales much larger than the cutoff scale for the PMF which is generated before the end of the BBN epoch, $k_\mathrm{FI}\sim 10^{-13} \mathrm{cm}^{-1}$.
In the case where $n_\mathrm{B}$ is less than -2.6 and $B_\lambda = 10$ nG, the $k_\mathrm{[max]}$ value is allowed to be larger than $10^{-13}$ cm $^{-1}$ (Fig.~\ref{fig5}). On the other hand, if $n_\mathrm{B}$ is larger than -2.6, the allowed value of $k_\mathrm{[max]}$ becomes smaller.
However the upper limits on $k_\mathrm{[max]}$ for very large $n_\mathrm{B}$ are smaller than the cosmic expansion rate, i.e., $H$ (the lower limits on $k_\mathrm{[max]}^{-1}$ are larger than the Horizon scale) before BBN,
and such a PMF can only be generated before the end of inflation.
Therefore, using this feature, the PL index of the PMF generated in the post-inflation epoch can be strongly constrained.
\subsection{\label{subsec:discusses2} Energy transfer from PL-PMF}
From Fig.~\ref{fig2}, it is confirmed that this BBN model including the effect of the PMF energy density cannot be a solution to the discrepancy between $^7$Li abundances produced in SBBN and those deduced from observations of MPHSs.  We consider a case where part of the PMF energy ($\rho_\mathrm{d}$) is dissipated and transferred to the radiation energies.
The final value of the photon energy density measured at an epoch of scale factor $a$ is then $\rho_\gamma(a) = \rho^{\mathrm{p}}_\gamma(a) + \rho_\mathrm{d}(a)$, where $\rho^{\mathrm{p}}_\gamma(a)$ is a primary photon energy density which does not include that transferred from the PMF by dissipation.
Since $\rho_\mathrm{d}/\rho_\gamma$ increases as a function of time from the time of the PMF generation, the following equation always holds:
$\rho_\mathrm{d}(a_\mathrm{BBN})/\rho_\gamma(a_\mathrm{BBN}) < \rho_\mathrm{d}(a_0)/\rho_\gamma(a_0)$.
This means that the baryon to photon ratio at the BBN epoch is expected to be more than the present value, and the $\eta$ value in the BBN epoch constrained from the WMAP 7yr data would be larger than the case of no energy transfer.

Here, we conservatively consider the most general case in which the
history of generation and dissipation of the PMF in the Universe is
completely unrestricted. The amplitude of the PMF in the BBN epoch and the
largeness of the PMF energy dissipation during a time from BBN to the
cosmological recombination should then be regarded as two independent
parameters. The effect of the PMF on the cosmic expansion rate increases
 the abundances of $^4$He, D and $^3$He, and decreases that of $^7$Li for
$\eta$ values around the WMAP 7yr estimation for the SBBN ($\eta_\mathrm{WMAP}$). The effect of dissipation, on
the other hand, increases abundances of $^4$He and $^7$Li and decreases
those of D and $^3$He (Figs.~\ref{fig1}~and~\ref{fig2}). Effects of the PMF and dissipation on the abundances
of D, $^3$He and $^7$Li work in opposite directions, while they both increase $^4$He abundance. One can,
therefore, understand that an upper limit on the strength of the PMF from $^4$He, i.e.,
$B_\mathrm{SI}^{Y_\mathrm{p}}$, can be derived by assuming that a PMF
dissipation does not occur.  In this regard, the case of Fig.~\ref{fig1}
corresponds to the maximum amplitude of the PMF allowed by the constraint
from $^4$He abundance for $\eta \geq \eta_\mathrm{WMAP}$. 
The minimum $^7$Li abundance in the BBN model with a free
parameter of $B_\mathrm{SI}^{\mathrm{X}}$ and $\eta \geq \eta_\mathrm{WMAP}$ is given by the
cross region of the solid line for $^7$Li/H and the vertical band (Fig.~\ref{fig1}), 
which is higher than the abundances in MPHSs.
We then conclude that the abundances of $^4$He and $^7$Li cannot be
simultaneously consistent with observations [Eqs.~(\ref{limited_Yp_AOS}) and (\ref{limited_7Li})],
even in the case where both the energy density and the dissipation of the PMF are taken into account.
\section{\label{sec:summary}Summary}
\begin{figure}
\includegraphics[width=0.5\textwidth]{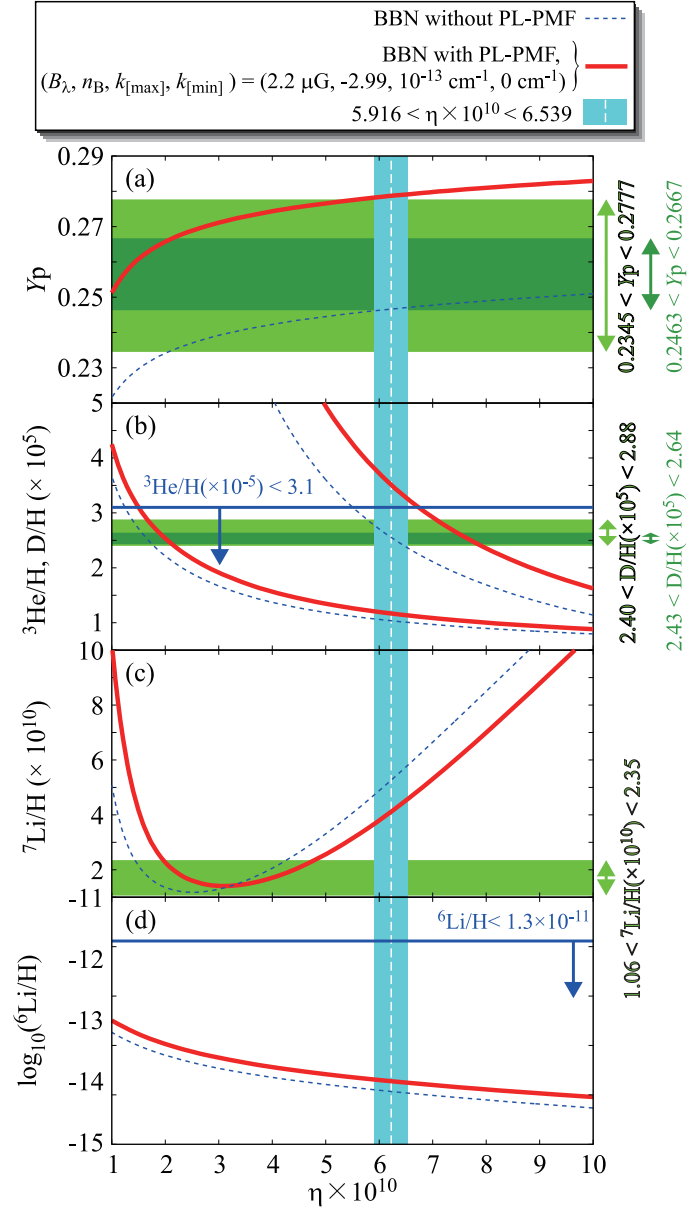}
\caption{\label{fig1}
Abundances of $Y_\mathrm{p}$(mass fraction of $^4$He), D/H, $^3$He/H, $^7$Li/H and $^6$Li/H as a function of $\eta$ in BBN models with and without the effect of the PL-PMF for the fixed value of $\lambda = 1$~Mpc.
The vertical painted band is the limit on the baryon to photon ratio from WMAP 7yr data \cite{2011ApJS..192...16L}. The white dashed line in the limited range of $\eta$ indicates the best value. 
The horizontal painted bands and lines with downward arrows indicate observational constraints and upper limits, respectively, on elemental abundances. 
The constraints on $^3$He and $^6$Li abundances are from Ref. \cite{2002Natur.415...54B} and Refs. \cite{2010IAUS..265...23S,2006ApJ...644..229A}, respectively.}
\end{figure}

\begin{figure}
\includegraphics[width=1.0\textwidth]{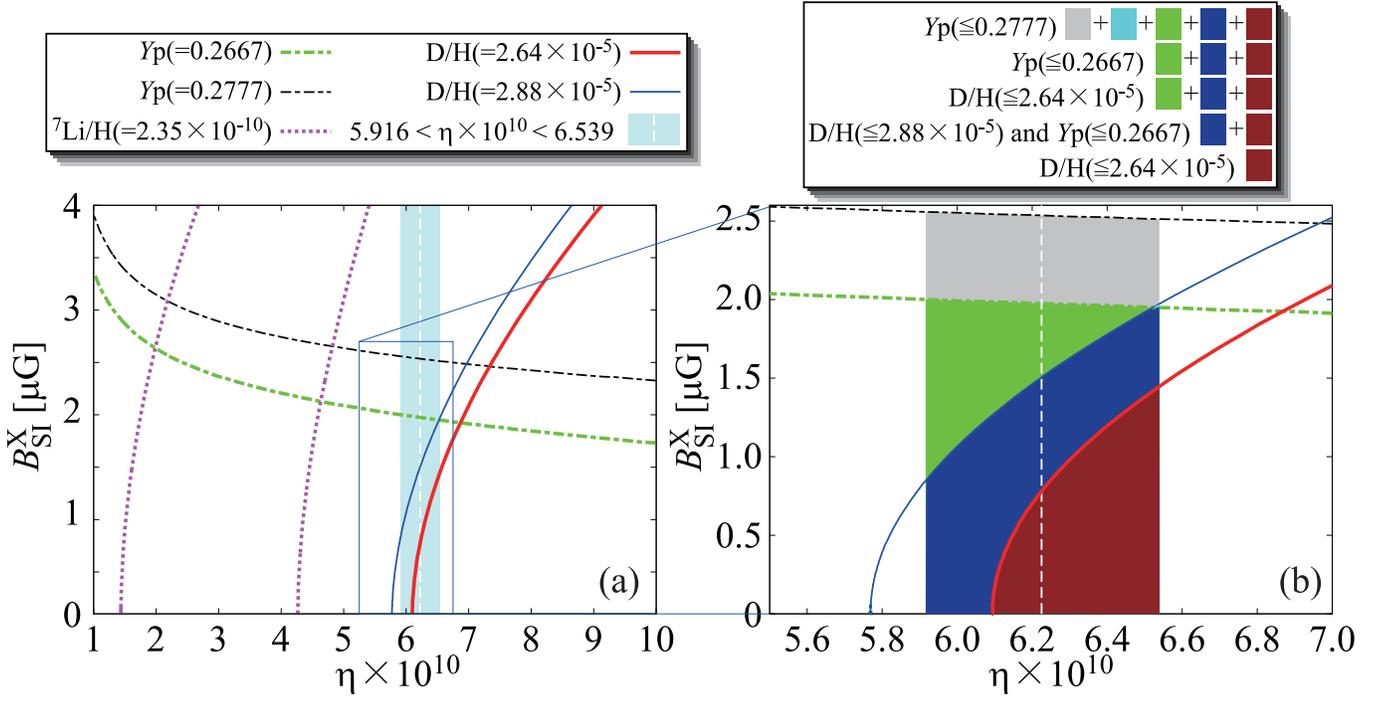}
\caption{\label{fig2}
Constraints on the field strength $B^\mathrm{X}_\mathrm{SI}$ derived from the consistency of the abundances of light elements, D, $^4$He and $^7$Li.
We use the following upper limits on the abundances: $Y_\mathrm{p}$ = 0.2667 (thick dot-dashed solid line) and 0.2777 (thin dote-dashed solid line), D/H = 2.88$\times 10^{-5}$ (thin solid line) and 2.64$\times 10^{-5}$ (thick solid line), and $^7$Li/H = 2.35$\times 10^{-10}$ (dotted line) [panel (a)]. The vertical painted band is the limit on the baryon to photon ratio from WMAP 7yr data \cite{2011ApJS..192...16L}. The white dashed line indicates the best value.  The right panel shows an enlarged view of allowed regions constrained from observations of CMB and elemental abundances. Painted regions in panel (b) are as indicated in the legend box at the top.} 
\end{figure}

\begin{figure}
\includegraphics[width=1.0\textwidth]{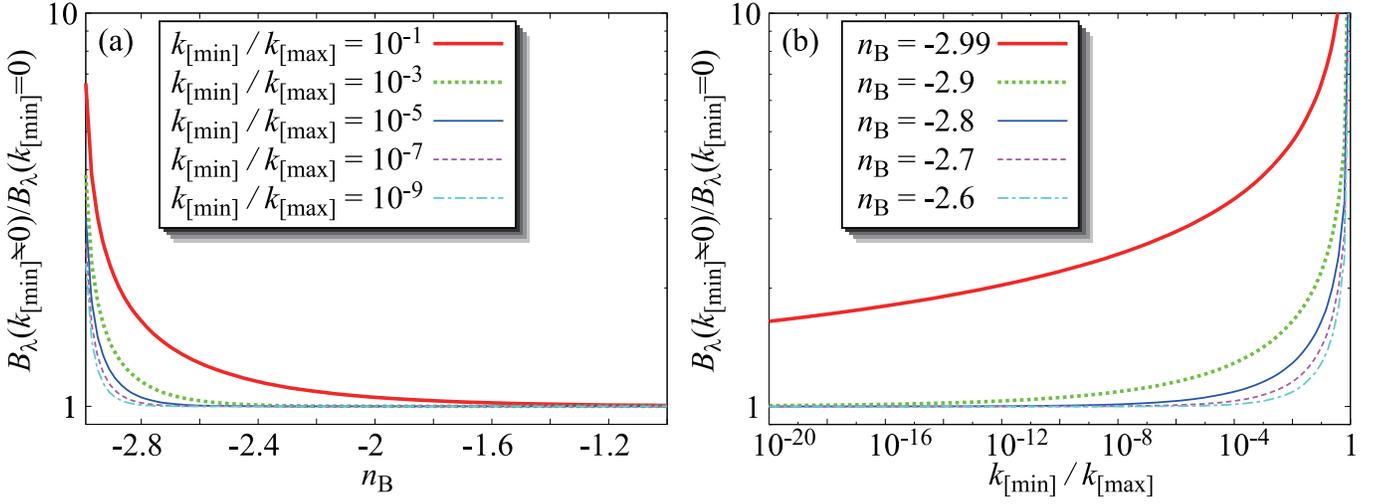}
\caption{\label{fig3}
Ratio of the magnetic field $B_\lambda(k_\mathrm{[min]}\ne 0)/ B_\lambda(k_\mathrm{[min]}=0)$ as a function of $n_\mathrm{B}$ [cf. Eqs.(\ref{eq:field_strength_for_elements}) and (\ref{upper_B_Yp_IT1})] for fixed ratios of wave number $k_\mathrm{[min]}/k_\mathrm{[max]}$ as shown in the legend box (a).  The right panel shows $B_\lambda(k_\mathrm{[min]}\ne 0)/ B_\lambda(k_\mathrm{[min]}=0)$ as a function of $k_\mathrm{[min]}/k_\mathrm{[max]}$ for fixed values of $n_\mathrm{B}$ (b).}
\end{figure}

\begin{figure}
\includegraphics[width=1.0\textwidth]{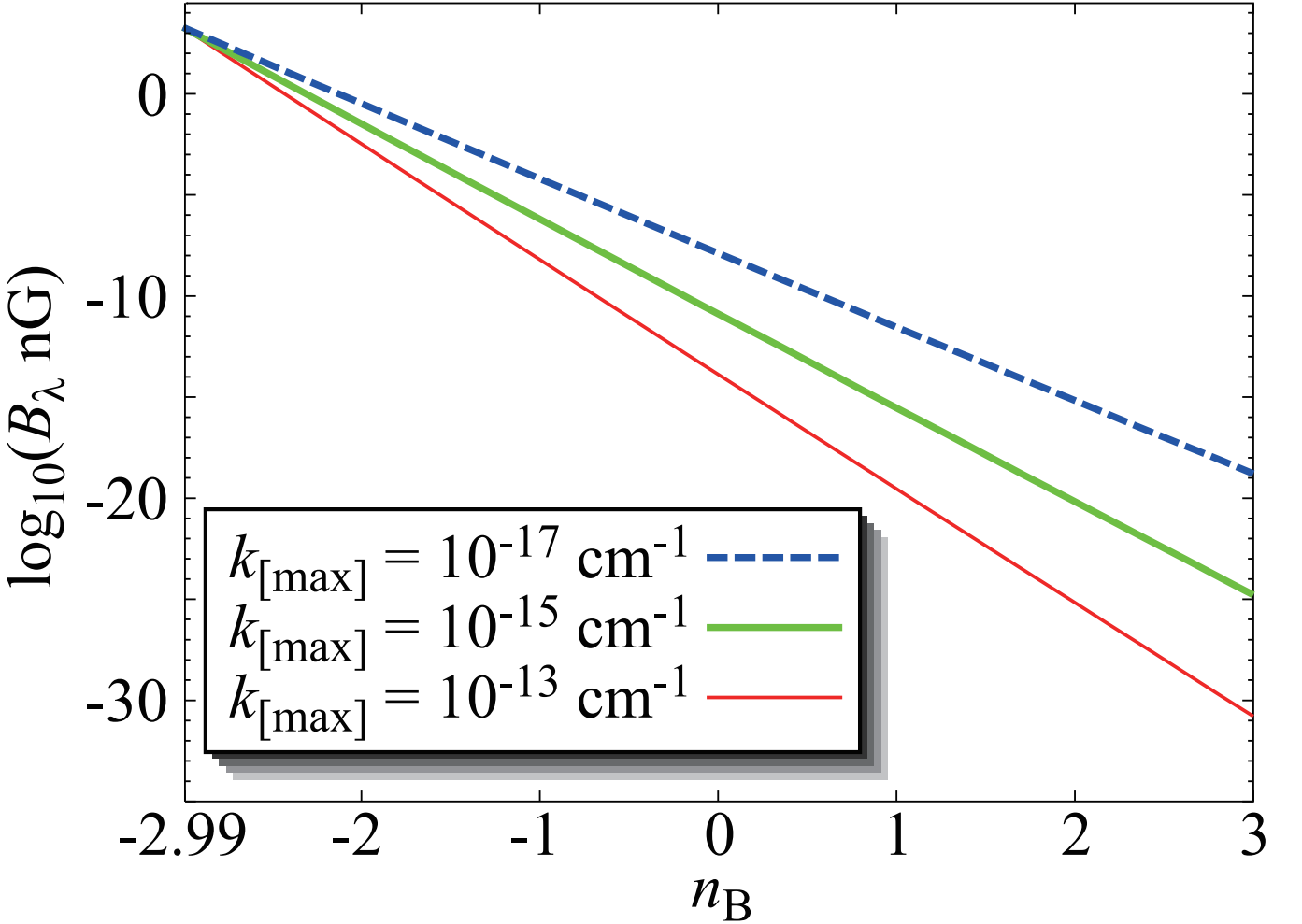}
\caption{\label{fig4}
Constraints on the field strength $B_\lambda $ as a function of $n_\mathrm{B}$ [cf. Eqs.(\ref{eq:field_strength_for_elements}) and (\ref{upper_B_Yp_IT1})] for fixed values of $\lambda = 1$Mpc and $k_\mathrm{[min]} =  0$.
The thin (red), thick (green), and dashed-thick (blue) lines show upper limits for cases of $k_\mathrm{[max]} = 10^{-13}$, $10^{-15}$, and $10^{-17}$, respectively.} 
\end{figure}

\begin{figure}
\includegraphics[width=1.0\textwidth]{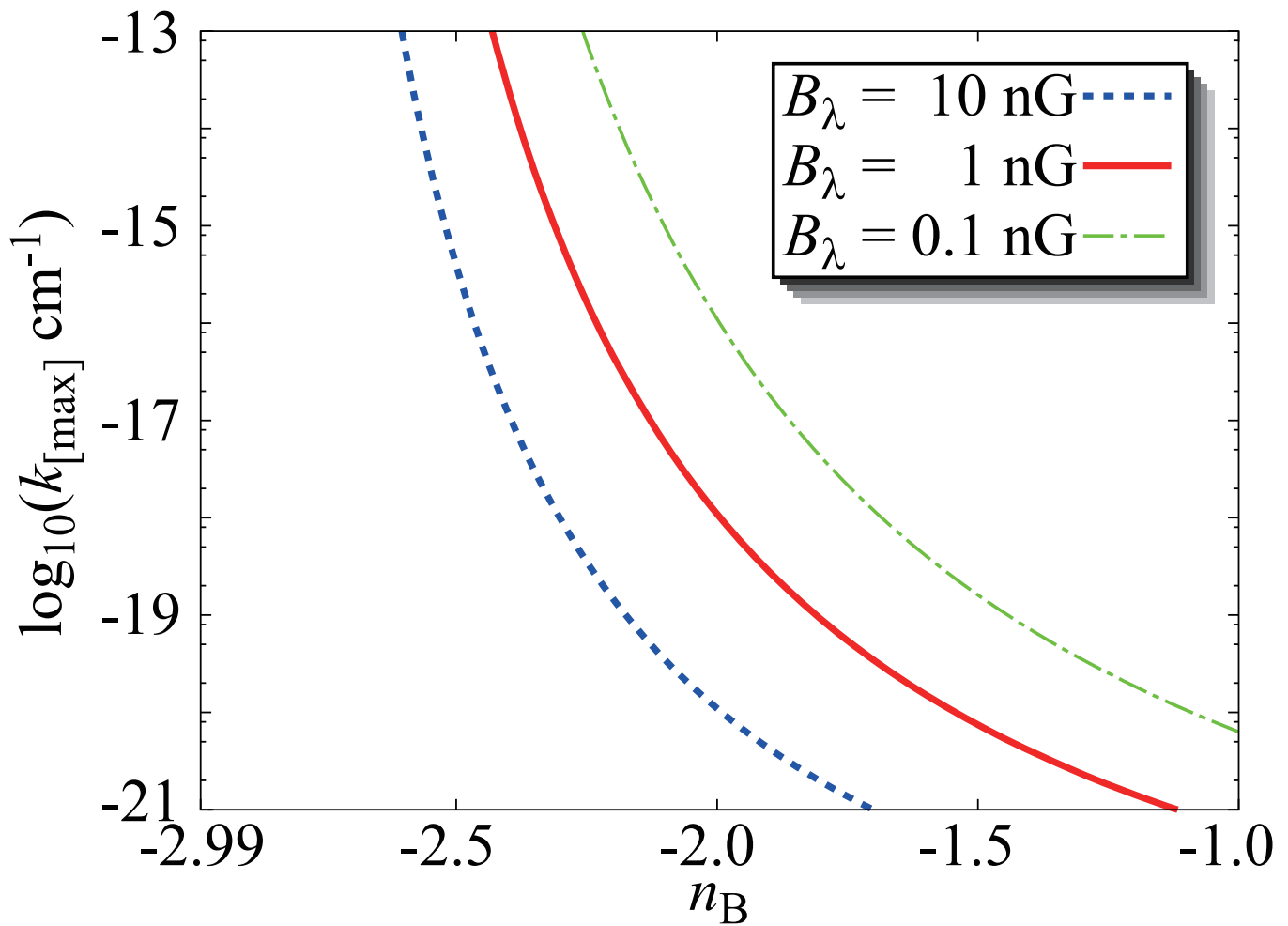}
\caption{\label{fig5}
Constraints on the maximum wave number $k_\mathrm{[max]}$ as a function of $n_\mathrm{B}$ [cf. Eqs.(\ref{eq:field_strength_for_elements}) and (\ref{upper_B_Yp_IT1})] for fixed values of $\lambda = 1$Mpc and $k_\mathrm{[min]} = 0$.  Results for various $B_\lambda$ values of 10 nG (dashed line), 1 nG (solid line) and 0.1 nG (dot-dashed line) are shown.
} 
\end{figure}

We showed quantitative and qualitative effects of a PMF on each abundance of a light element using a BBN model taking account of the PMF with two parameters of the SI strengths of PMFs and the baryon to photon ratio $\eta$ (Figs. \ref{fig1} and \ref{fig2}).

We assumed a PMF characterized by the power spectrum of a PL in wave number.  Parameters of this PL-PMF are a field amplitude $B_\lambda$, a PL index $n_\mathrm{B}$, and maximum and minimum scales at a generation epoch $k_\mathrm{[max]}$ and $k_\mathrm{[min]}$, respectively.  We then showed a relation between PL-PMF parameters and the SI strength of the PMF which is the parameter constrained from observations of light element abundances (Sec. \ref{subsec:models_ED_PLPMF}).  

We then calculated BBN taking into account recent updates on reaction rates and the effect of the PMF, and showed results of the abundances in the example case of ($B_\mathrm{1~Mpc}, n_{\rm B}, k_{\rm [max]}, k_{\rm [min]})=(2.2~\mu$G, $-2.99$, $10^{-13}$ cm$^{-1}$, $0$ cm$^{-1}$) as a function of $\eta$.  It was observed that the critical $\eta$ value, i.e., $\eta_{\rm crit}$, at which $^7$Li and $^7$Be are produced in the same abundance at the BBN epoch, moves by PMF effects. 
The value $\eta_{\rm crit}= 3.00 \times10^{-10}$ for SBBN is enhanced up to $\eta_{\rm crit}= 3.71 \times10^{-10}$ for the example case.  In addition, the minimum of the final $^7$Li abundance, located at the valley of the $^7$Li/H vs. $\eta$ curve moves from $^7$Li/H$_{\rm min}=1.18\times10^{-10}$ at $\eta=2.50 \times 10^{-10}$ for the SBBN to $^7$Li/H$_{\rm min}=1.41\times10^{-10}$ at $\eta=3.10 \times 10^{-10}$ for BBN with the PL-PMF (Sec. \ref{subsec:BBNwithPMF}).

We investigated constraints on parameters of the PL-PMF from the abundances of light elements up to Li produced in BBN.  
We showed respective and combined constraints on the parameters. 
As a result, we obtained upper limits on the SI strength of the PMF, i.e., $B^{Y_\mathrm{p} + D}_\mathrm{SI} \le 1.45 ~ - ~ 1.95 \mu$ G from abundances of $^4$He and D for the 2 $\sigma$ region of $\eta$ values by the WMAP 7yr data (Sec.~\ref{subsec:discussions2}). 

In Sec.IV, we discussed the degeneracy of the PL-PMF parameters in effects on BBN and showed some possibility of constraining models of PMF generations by combining constraints on the SI field strength from light element abundances to the relation between PL-PMF parameters and the SI field strength (Sec. \ref{subsec:discussions1}). 

In addition, we consider a general case in which the existence and
energy dissipation of the PMF are allowed.
 Based on our result of the BBN calculation (Figs. \ref{fig1} and \ref{fig2}), it was found that an upper limit on the strength of the PMF can be derived from a constraint on $^4$He abundance.  A lower limit on $^7$Li abundance is also derived, and it is significantly higher than those observed in metal-poor stars.  We then conclude that it is impossible to solve the $^7$Li problem by the PMF energy density, even if we consider that part of the PMF energy is dissipated and transferred to the radiation energies  (Sec.~\ref{subsec:discusses2}).

In the near future, cosmological observations of the CMB and large scale structures will make it possible to constrain the PL-PMF parameters more tightly. If we combine such results and our constraint on the parameters through the study of BBN, we will not only limit the PL-PMF parameters but also the generation models of the PMF more precisely.

\begin{acknowledgments}
We are grateful to Professor G. J. Mathews and Professor T. Kajino for valuable discussions. This work was partly supported by Grant-in-Aid for JSPS Fellows No.21.6817 (M.K.).
\end{acknowledgments}
\bibliographystyle{apsrev}

\begin{thebibliography}{87}
\bibitem{f_note_1}
Avoiding divergence of Eq. (\ref{eq:field_strength_for_elements}), we deal with $k_\mathrm{[min]}/k_\mathrm{[max]} \le 0.99$ and $n_\mathrm{B} \ge -2.99$ in this paper.
\end{thebibliography}

\end{document}